\documentclass{elsart}
\usepackage{amssymb}
\usepackage{graphicx}  

\newcommand{\bbox}[1]{\mbox{\boldmath${#1}$\unboldmath}}

\begin{document}

\begin{frontmatter}
\title{Large-Scale Self-Consistent Nuclear Mass Calculations}
\author[label1,label2,label3,label4]{M.V.~Stoitsov},
\author[label1,label2,label5]{J.~Dobaczewski},
\author[label1,label2,label5]{W.~Nazarewicz},
\author[label1,label2,label6]{P.~Borycki}
\address[label1]{Department of Physics \& Astronomy, University of Tennessee, Knoxville, Tennessee 37996, USA}
\address[label2]{Physics Division,  Oak Ridge National Laboratory, P.O. Box 2008, Oak Ridge, Tennessee 37831, USA}
\address[label3]{Joint Institute for Heavy-Ion Research, Oak Ridge, Tennessee 37831, USA}
\address[label4]{Institute of Nuclear Research  and Nuclear Energy, Bulgarian Academy of Sciences, Sofia-1784, Bulgaria}
\address[label5]{Institute of Theoretical Physics, Warsaw University, ul. Ho\.za 69, 00-681 Warsaw, Poland}
\address[label6]{Institute of  Physics, Warsaw University of Technology, ul. Koszykowa 75, 00-662 Warsaw, Poland}

\begin{abstract}
The program of systematic large-scale self-consistent nuclear mass
calculations that is
based on the nuclear density functional theory
represents a rich scientific agenda that is closely aligned
with the main research directions in modern nuclear structure and astrophysics,
especially  the radioactive nuclear beam  physics. The quest for the
microscopic understanding of the phenomenon of nuclear binding
represents, in fact, a number of fundamental  and  crucial  questions of
the quantum  many-body problem, including the proper treatment of
correlations and dynamics in the presence of symmetry breaking. Recent
advances and open problems in the field of nuclear mass calculations are
presented and discussed.
\end{abstract}

\end{frontmatter}

\section{Introduction}

The study of nuclei far from stability  is an increasingly important
part of a nuclear physics portfolio \cite{NSAC,NSAC1,NUPECC}. As
radioactive beams gradually expand the borders of the nuclear landscape,
theoretical modeling of the nucleus is changing in significant ways. The
crucial  question for the field \cite{NSAC1}, namely ``What binds protons and
neutrons into stable nuclei and rare isotopes?," nicely underlines  this
point: indeed,
 the data on rare isotopes with the large neutron-to-proton imbalance
indicate that there are many gaps in our present understanding.

Short-lived exotic nuclei offer
unique tests of those aspects of the nuclear  theory
that depend on neutron excess~\cite{[Dob98c],RIAT}. The major
challenge is to predict or describe in detail exotic new properties
of nuclei far from the stability valley, and to explain the
origins of these properties. New ideas and progress in computer
technology have allowed nuclear theorists to understand bits and
pieces of nuclear structure quantitatively.

The new experimental developments inevitably require safe and reliable
theoretical predictions of nuclear properties throughout the whole
nuclear chart in   two   main directions: (i) along the
isospin axis, i.e., going outwards from the beta stability line to the neutron
and proton drip lines, and (ii) towards the uncharted territory of
super-heavy elements at the limit of mass and charge.
The tool of choice is the nuclear density
functional theory (DFT) based on the self-consistent
Hartree-Fock-Bogoliubov (HFB) method. The key component is
the universal energy density functional, which will be able to describe
properties of finite nuclei as well as extended asymmetric nucleonic
matter. The development  of such a universal functional,
including dynamical effects and symmetry restoration,
 is one of  the
main goals of the field.

By employing various criteria (agreement with
measured masses, radii, low-lying excited states, giant vibrations,
rotational properties, and other global nuclear characteristics), one
aims at  adjusting the coupling constants of the functional. By finding
correlations between parameters, one hopes to reduce their number
and to understand physical reasons why different parametrizations
yield similar results. One may also want to
expand the parametrizations to cover aspects dictated by physics
arguments and/or motivations coming from the effective field theory
and QCD. The main challenges in this quest have been nicely
summarized through five questions
\cite{INT05}:
\begin{itemize}
\item What is the form of the nuclear energy density functional?
\item What are the constraints on the nuclear energy density functional?
\item What is the form of the pairing functional?
\item How to account for quantum correlations and symmetry-breaking effects?
\item How to optimize computational techniques and error analysis?
\end{itemize}

The aim of this paper is to briefly  review the present state of
the large-scale microscopic nuclear mass calculations and to discuss
 improvements needed. Section~\ref{sect1}
introduces   the DFT and Skyrme-HFB method. Some
 details concerning global 
 mass calculations are given  in
Sec.~\ref{sect3}. The long-term program
is outlined in Sec.~\ref{sect5}. Finally, the
summary is given in
Sec.~\ref{sect6}.

\section{Nuclear Energy Density Functional}
\label{sect1}

A theoretical framework aiming at the microscopic description of nuclear
masses and capable of extrapolating into an unknown territory must fulfill
several strict requirements. First, it must be general enough to be
confidently applied to a region of the nuclear landscape whose
properties are largely unknown. Second, it should be capable of
handling symmetry-breaking effects resulting in a large variety of
intrinsic nuclear deformations. Thirdly,
it should describe finite nuclei and the bulk nuclear matter.
Finally,  in addition to observables, the method should provide
associated error bars.

These requirements are met by the DFT in
the formulation of Kohn and Sham \cite{ks65}.
The main ingredient of the non-relativistic nuclear
DFT \cite{[Ben03]}
(for relativistic  nuclear DFT, see Ref.~\cite{lal04}) is the energy density functional
that depends on densities and
currents representing distributions of nucleonic matter, spins,
momentum, and kinetic energy, as well as their derivatives (gradient
terms). Standard Skyrme functionals employed in self-consistent
mean-field calculations are parametrized by means of about ten coupling
constants that are adjusted to basic properties of nuclear matter (e.g.,
saturation density, binding energy per nucleon) and to selected data on
magic nuclei. The functionals are augmented by the pairing term which
describes nuclear superfluidity \cite{[Per03]}. When not corrected by
additional
phenomenological terms, standard functionals reproduce total binding
energies with an rms error of the order of
2 to 4 MeV \cite{[Pat99],[Lun03],[Gor05]}. However, they
have been successfully tested over the whole nuclear chart to a broad
range of phenomena, and usually perform quite well when applied to
energy differences, radii, and nuclear moments and deformations
\cite{[Ben03]}.

Historically, the first
nuclear energy density functionals appeared in the context of
Hartree-Fock (HF) or HFB methods and zero-range
interactions such as the Skyrme force.
However, it was  realized afterwards that -- in the spirit of
the DFT --
an effective interaction could be secondary
to the functional, i.e., it is the  density functional that defines
the force. This is the strategy that we are going to follow.

\subsection{The Densities}

The main ingredients of the nuclear DFT are the local nucleonic densities.
Following the standard definitions \cite{[Eng75],[Per03]}, one considers
local particle-hole (p-h) densities:
particle $\rho (\bbox{r})$,
kinetic $ \tau (\bbox{r})$,
spin $ {\bbox{s}}_k (\bbox{r})$,
spin-kinetic $ {\bbox{T}}_k (\bbox{r})$,
current $ {\bbox{j}}_k (\bbox{r})$,
tensor-kinetic $ {\bbox{F}}_{k} (\bbox{r})$,
spin-current ${\mathsf{J}}_{kl}(\bbox{r})$,
 as well as the corresponding
local particle-particle (p-p; or pairing) densities:
$\tilde{\rho}(\bbox{r})$,
$  \tilde{\tau} (\bbox{r})$,
$ {\tilde{\bbox{s}}}_k (\bbox{r})$,
$ {\tilde{\bbox{T}}}_k (\bbox{r})$,
$ {\tilde{\bbox{j}}}_k (\bbox{r})$,
$ {\tilde{\bbox{F}}}_{k} (\bbox{r})$,
and ${\tilde{\mathsf{J}}}_{kl}(\bbox{r})$.

The local p-h and p-p densities
are defined by the spin-dependent one-body density matrices:
\begin{equation}
\begin{array}{rcl}
\rho (\bbox{r}\sigma ,\bbox{r^{\prime }}\sigma^{\prime })
&=&\displaystyle\frac{1}{2} \rho (\bbox{r},\bbox{r}^{\prime
})\delta_{\sigma \sigma^{\prime }} + \frac{1}{2} \sum_{i}(\sigma
|\sigma_{i}|\sigma^{\prime })\rho_{i}(\bbox{r},\bbox{r}
^{\prime }) ,   \\ && \\
\tilde{\rho}(\bbox{r}\sigma ,\bbox{r^{\prime }}\sigma^{\prime })
&=&\displaystyle\frac{1}{2 } \tilde{\rho}(\bbox{r},\bbox{r}^{\prime
})\delta _{\sigma \sigma^{\prime }}+\frac{1}{2} \sum_{i}(\sigma
|\sigma_{i}|\sigma^{\prime })\tilde{\rho}_{i}( \bbox{r},\bbox{r}^{\prime }) .
\label{densitieemp}
\end{array}
\end{equation}

For instance,
\begin{equation}
\begin{array}{rcl}
\rho (\bbox{r})  =  \rho (\bbox{r},\bbox{r}),~~~ \tau (\bbox{r})  &=&
\left. \nabla_{\bbox{r}}\nabla_{\bbox{r}^{\prime }}\rho (\bbox{r},\bbox{r}^{\prime })
\right|_{\bbox{r}^{\prime }\bbox{=r}}, ~~~
\tilde{\rho}(\bbox{r})  =  \tilde{\rho}(\bbox{r},\bbox{r}), \\
~ &  &  \\
\mathsf{J}_{ij}(\bbox{r}) & = & \frac{1}{2i}\left. \left( \nabla
_{i}-\nabla_{i}^{\prime }\right) \rho_{j}(\bbox{r},\bbox{r}^{\prime
})\right|_{\bbox{r}^{\prime }\bbox{=r}}\;.
\end{array}
\label{densities}
\end{equation}
Since the nuclear DFT deals with two kinds of nucleons, the
isospin degree of freedom has to be introduced and
the isoscalar and isovector densities have to be considered\cite{[Per03]}.

\subsection{The Energy Density Functional}

The energy density functional has the form
\begin{equation}
E[\rho,\tilde{\rho}]=\int d^3\bbox{r}~{ \mathcal{H} }(\bbox{r}),
\label{shfb}
\end{equation}
where energy density $\mathcal{H}(\bbox{r})$ is
usually written as a sum of the
p-h energy density $H(\bbox{r})$ and the p-p
energy density $\displaystyle \tilde{H}(\bbox{r})$.
According to the DFT, there exists
a nuclear  universal energy functional that yields the exact
binding energy of the  nuclear system.
This functional includes, in principle, all
many-nucleon correlations.

The actual form of the nuclear energy functional is unknown. The
strategy adopted by many practitioners is to build a functional around
that generated by the Skyrme interaction. The most general form of the
energy density functional that is quadratic in local densities and
preserves the basic symmetries of the strong interaction, such as
time-reversal symmetry, charge invariance, and proton-neutron symmetry,
has been proposed in Ref.~\cite{[Per03]}.
In practical applications, however, simplified forms of this functional
have been used. For instance, one particular representation
of the energy functional for the ground states  of
even-even nuclei can be written
as::
  \begin{equation}
\begin{array}{rll}
 H(\bbox{r}) &=&  \frac{\hbar^{2}}{2M}~ \tau
+\frac{1}{2}t_{0}\left[ \left( 1+\frac{1}{2}x_{0}\right)
\rho^{2}\right. -\left( \frac{1}{2}+x_{0}\right)
\sum\limits_{q}\left.
\rho_{q}^{2} \right] \\
& + & \frac{1}{4}t_{1}\left[ \left( 1+\frac{1}{2}x_{1}\right) \rho
\left( \tau -\frac{3}{4}\left. \Delta \rho \right) \right. \right]
-\left(\frac{1}{2}+x_{1}\right) \sum\limits_{q}\left. \rho
_{q}\left( \tau_{q}-\frac{3}{4}\Delta \rho_{q}\right) \right] \\
& + & \frac{1}{4}t_{2}\left[ \left( 1+\frac{1}{2}x_{2}\right) \rho
\left( \tau +\frac{1}{4}\Delta \rho \right) \right.  +\left(
\frac{1}{2}+x_{2}\right) \sum\limits_{q}\left. \rho
_{q}\left( \tau_{q}+\frac{1}{4}\Delta \rho_{q}\right) \right] \\
& + & \frac{1}{12}t_{3}\rho^{\alpha }\left[ \left( 1+\frac{1}{2}
x_{3}\right) \rho^{2}\right. -\left( x_{3}+\frac{1}{2}\right)
\sum\limits_{q}\left. \rho_{q}^{2}\right] - \frac{1}{8}\left(
t_{1}x_{1}+t_{2}x_{2}\right) \sum\limits_{ij}{\mathsf{J}}_{ij}^{2}
\\
& + &  \frac{1}{8}\left( t_{1}-t_{2}\right) \sum\limits_{q,ij}{\mathsf{J}}_{q,ij}^{2} -
\frac{1}{2}W_{0}\sum\limits_{ijk}\varepsilon_{ijk}\left[ \rho
{\bbox{\nabla}}_{k}{\mathsf{J}}_{ij}\right. +\sum\limits_{q}\left.
\rho_{q}{\bbox{\nabla}}_{k}{\mathsf{J}}_{q,ij}\right] \\
& + &  {H}^C(\bbox{r})
\end{array}
\label{skyrmeph}
\end{equation}
and
\begin{equation}
\displaystyle \tilde{H}(\bbox{r}) = \frac{1}{2} V_{0}
\left[1-V_1\left(\frac{\rho}{\rho_0}\right)^\gamma~
\right]\sum\limits_{q}\tilde{\rho}_{q}^{2}~, \label{edhppd}
\end{equation}
where $q$  labels the neutron
($q=n$) or proton ($q=p$) densities and the
quantities  which do not carry  index $q$  are
the  isoscalar densities (sums of proton and neutron densities;
e.g., $\rho \equiv \rho(\bbox{r})=\rho_n(\bbox{r})+\rho_p(\bbox{r})$).

The p-p energy functional (\ref{edhppd}) corresponds to a density-dependent
delta interaction. Usually,
$\gamma=1$, $\rho_0$=0.16 fm$^{-3}$, and
$V_1$=0, 1, or 1/2 for
volume-, surface-, or mixed-type pairing. In Eq.~(\ref{skyrmeph}),
$H^C(\bbox{r})$
stands for the Coulomb energy density with the exchange term treated
in the Slater approximation.

As seen from Eqs.~(\ref{skyrmeph}) and (\ref{edhppd}),
typical Skyrme
density functionals include about 14 unknown parameters. Some of
them are usually adjusted to reproduce the basic
properties of the  infinite nuclear matter while the remaining coupling
constants  are fitted to known nuclear masses, radii, and other
measured properties.

\subsection{Variational Equations}

By varying the energy functional (\ref{shfb}) with respect to the
density matrices $\rho $ and
$\tilde{\rho}$ one arrives at the HFB equations:
\begin{equation}\label{eq143}
\left(\begin{array}{cc}
h-\lambda &
\tilde h \\
\tilde h & - h+\lambda
\end{array}\right)
\left(\begin{array}{c}
U \\
V
\end{array}\right) = E
\left(\begin{array}{c}
 U \\
 V
\end{array}\right),
\end{equation}
where
$U=U(E,\bbox{r}\sigma)$, $V=V(E,\bbox{r}\sigma)$
are the HFB wave functions, and $h$ and $\tilde h$ are
the local particle and pairing mean-field Hamiltonians.

The HFB equations (\ref{eq143}), also called the  Bogoliubov de Gennes
equations by condensed matter physicists, are the generalized Kohn Sham
equations of the DFT. It is worth noting that - in its  original
formulation \cite{Oli88} - the DFT formalism implicitly includes the
full correlation functional. In most nuclear applications, however, the
correlation corrections are added afterwards. Those corrections usually
include the following terms: the  center-of-mass correction, rotational
correction associated with the spontaneous breaking of rotational
symmetry, vibrational correction (quantum zero-point vibrational
fluctuations), particle-number correction due to the broken gauge
invariance, as well as other terms.

The spectrum of quasi-particle energies $E$ is continuous for
$|E|$$>$$-\lambda$ and discrete for $|E|$$<$$-\lambda$.
However, when
solving the HFB equations on a coordinate-space lattice of points
or by expanding quasi-particle
wave functions in a finite basis, the quasi-particle spectrum
is discretized  and  one can use the notation $V_k(\bbox{r}\sigma)=V(E_k,\bbox{r}\sigma)$
and $U_k(\bbox{r}\sigma) =U(E_k,\bbox{r}\sigma)$.
Since for $E_k$$>$0 and $\lambda$$<$0 the lower
components $V_k(\bbox{r}\sigma)$ are localized functions of $\bbox{r}$,
the density matrices,
\begin{eqnarray}
\rho(\bbox{r}\sigma,\bbox{r}'\sigma') &=& \sum_k  V_k (\bbox{r} \sigma
) V_k^*(\bbox{r}'\sigma'),~~~ \tilde\rho(\bbox{r}\sigma,\bbox{r}'\sigma')
= - \sum_k  V_k(\bbox{r} \sigma ) U_k^*(\bbox{r}'\sigma'),
,   \label{densitt}
\end{eqnarray}
are always localized.
The norms $N_k$ of the lower components define the total number
of particles
\begin{equation}\label{norms}
N_k  = \sum_{\sigma} \int d^3\bbox{r}|V_k(\bbox{r} \sigma )|^2,~~~
\label{Ntot} N= \sum_{k}N_k =\int d^3\bbox{r}~\rho(\bbox{r}).
\end{equation}

For spherical nuclei, the self-consistent HFB equations are best
solved in the coordinate space where they form a set of 1D radial
differential equations \cite{[Dob84],[Dob96]}. In the case of
deformed nuclei, however, the solution of deformed HFB equations in
coordinate space is a difficult and time-consuming task. For axial nuclei,
the corresponding 2D differential equations can be solved by using
the basis-spline methods (see, e.g., Ref.~\cite{[Ter03]}).
For triaxial nuclei, 3D solutions in a restricted space are
possible by using the so-called two-basis method \cite{[Gal94]}.

\section{Large-Scale Microscopic Nuclear Mass Calculations}
\label{sect3}

The large-scale microscopic nuclear mass calculations, such as those of
Refs.\ \cite{[Gor05],[Gor03],[Sam04],[Sto03]}, typically require
that
the variational equations are repeatedly sol\-ved
for thousands of nuclei. For
example, when adjusting the parameters of the energy density functional
to  measured masses, one has to calculate ground-state configurations  of
around two-thousand nuclei  many times during the fitting process.
\begin{figure}[htb]
\includegraphics[width=0.9\textwidth]{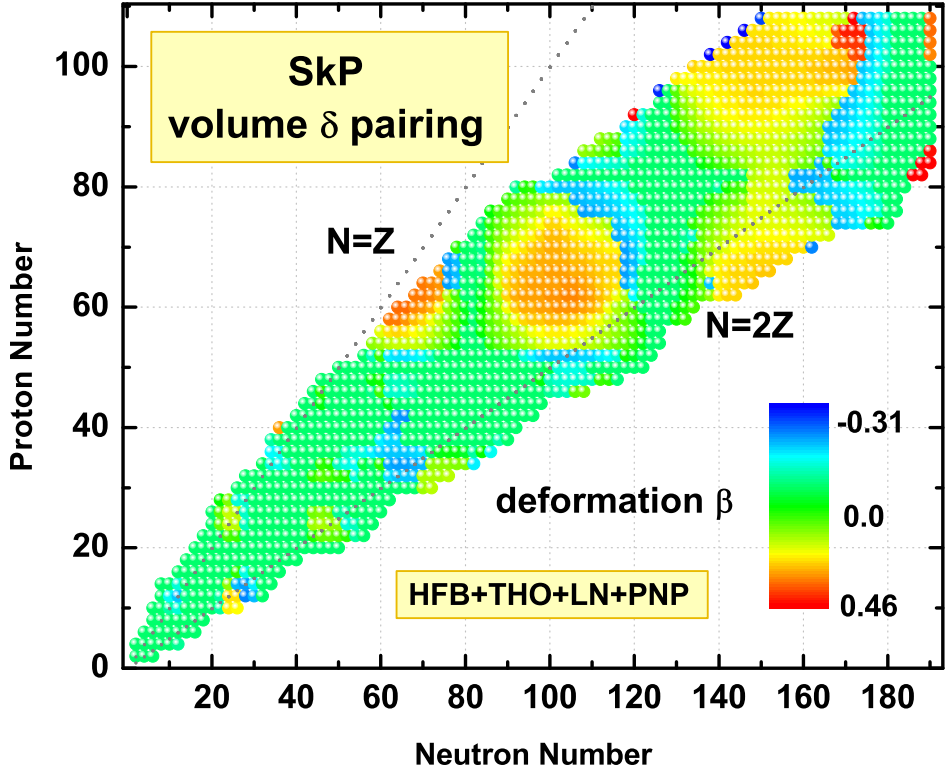}
\protect\caption{\label{mass1}Quadrupole deformations $\beta$ for all
even-even particle-bound nuclei  calculated with the
SkP energy
density functional \cite{[Dob84]}
in the p-h channel and the volume delta pairing using
the deformed HFB+THO code with 20 THO shells.}
\end{figure}
Actually, the situation is even more complicated, as several independent
calculations have to be carried out for a given nucleus  to find the
ground-state energy of the system among several coexisting local minima.
Furthermore, if odd-$A$ and odd-odd nuclei are considered during the
fitting process, many one-quasiparticle and two-quasiparticle states
have to be considered to find the actual ground state. Finally, when the
functional has been established, properties of around ten-thousand
particle-bound nuclei throughout the nuclear chart can be computed. All
in all, fitting a functional and preparing a mass table  is a
challenging computational problem that requires highly optimized
numerical codes and excellent  utilization of modern multiprocessor
computer resources.

\begin{figure}[htb]
 \includegraphics[width=0.9\textwidth]{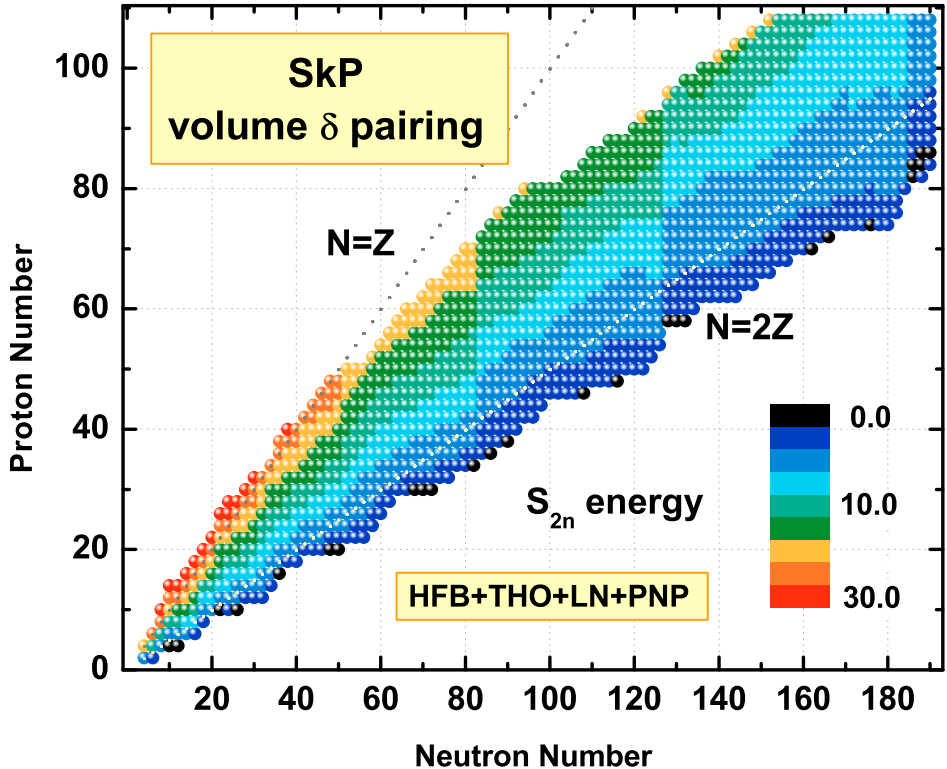}
\protect\caption{\label{mass2}Similar to  Fig.~\protect\ref{mass1}
except for two-neutron separation energies.}
\end{figure}
Our group has been
laying out theoretical foundations
and constructing computational tools to tackle this ambitious task.
We utilize  a fast HFBRAD code for spherical HFB calculations \cite{karim},
 which takes no more than 10 CPU minutes per nucleus on
an Intel Xeon 2.8 GHz processor, as well as the HFBTHO code for
axially deformed HFB calculations \cite{[Dob04a],hfbtho} with
acceptable processor speed -  averaging to about 1 CPU hour per nucleus.

The large-scale mass calculations based on the HFB+THO code,
extended with a minimal MPI communication in order to run in a
parallel regime across the nodes of the multiprocessor computer, are
illustrated in Figs.~\ref{mass1} and \ref{mass2}, which display,
respectively,
calculated charts of nuclear deformations  and two-neutron separation
energies for particle-bound  even-even nuclei.
We used the SkP energy functional \cite{[Dob84]}, which has a general
form given by Eqs.~(\ref{skyrmeph}) and (\ref{edhppd}).
Our load-balancing routine, which scales the problem to 200 processors,
allows us to perform these calculations in a single 24 wall-clock
hour run on a 4~Tflop machine Cheetah at ORNL (1~Tflop=$1 \times
10^{12}$ floation-point operations/sec) \cite{[Sto03]}.
For the details of the Skyrme-HFB deformed nuclear mass table
with SLy4 and SkP energy density functionals, see Ref.~\cite{masstable}.
In the following,
some particular aspects of our Skyrme-DFT
calculations are briefly discussed.

\subsection{Transformed Harmonic Oscillator Basis}

Going away from the beta stability valley
towards particle drip lines, the
Fermi energy becomes very small and the nucleonic densities and fields
acquire large spatial extensions due to the coupling to the particle
continuum. In this region of weakly bound nuclei, the asymptotic
behavior of nuclear densities has an effect on  nuclear properties.
Consequently, when performing calculations for drip-line systems,
it is important
to have a firm grasp on physics at large distances.
The recently developed HFB-THO technique
based on  the transformed harmonic oscillator  (THO) method
\cite{[Sto98],[Sto99],[Sto03]} is very helpful in this respect:
it is fast, efficient, and easy to implement.

\begin{figure}[htb]
\includegraphics[width=0.9\textwidth]{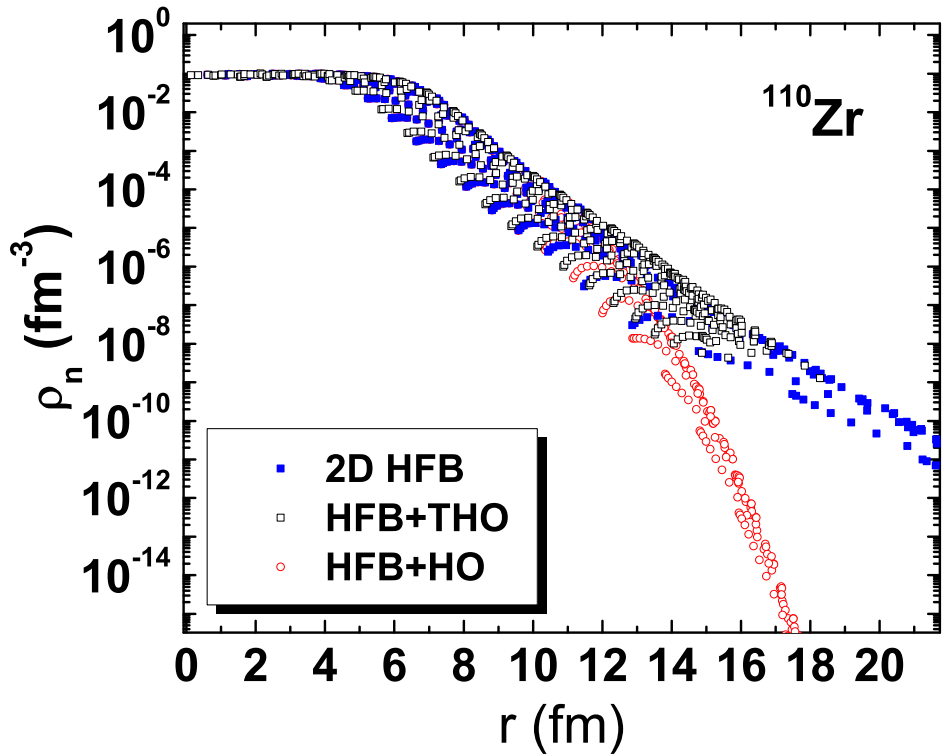}
\protect\caption{\label{densit}Comparison of the neutron densities
(in logarithmic scale) calculated for the deformed nucleus
$^{110}$Zn using  coordinate-space 2D calculations
(solid squares) with the configurational calculations based on
THO (open squares) and HO (open circles) basis \cite{[Umar]}.
Each point corresponds to
one Gauss-integration node in the $z-\rho$ plane, and the results
are plotted as functions of the distance from the origin,
$r=\sqrt{z^2+\rho^2}$.}
\end{figure}
Figure~\ref{densit} shows the neutron density
of  the deformed nucleus
$^{110}$Zn  obtained
in two
configurational calculations based on  expansions in the harmonic
oscillator (HO) and THO bases
\cite{[Sto98],[Sto99],[Sto03]} compared to  full-fledged
2D coordinate-space calculations \cite{[Obe03],[Umar]} with the box
boundary conditions.
Every point in the figure
corresponds to the value of the neutron density at a given
Gauss-integration node in the $z-\rho$ plane. Since the nucleus is
deformed, and there are always several nodes near a sphere of the
same radius $r = \sqrt{z^2 + \rho^2}$, there can be seen some
scatter of points corresponding to different densities in different
directions.
While the significant deviation from the correct
decaying  behavior is  seen in the HO results, the THO expansion
agrees very well with   the  deformed coordinate-space  method.
Other promising techniques that can alternatively be used in this
context are the
Gaussian-expansion basis method \cite{[GEM]} and the Berggren
expansion method \cite{[GSM]}.

\subsection{Regularization of the Contact Pairing Interaction}

When employing contact pairing interactions such as the density-dependent
delta force resulting in the pairing functional
(\ref{edhppd}), one has to apply a cut-off procedure and use a finite
space of single-particle states \cite{[Dob84]}.
When this space increases, the
pairing energy diverges for any strength of the interaction;
therefore, one has to readjust the pairing strength for each size of the
single-particle space \cite{[Dob96]}.
Such  renormalization procedure
is performed in the spirit of the effective
field theory, whereupon contact interactions are used to describe
low-energy phenomena while the coupling constants are readjusted for
any given energy cut-off to take into account neglected high-energy
effects. It has been shown that by carrying out renormalization
for each value of the cut-off energy, one
practically eliminates the dependence of the HFB results on the
size of the single-particle space.

\begin{figure}[htb]
\includegraphics[width=1.0\textwidth]{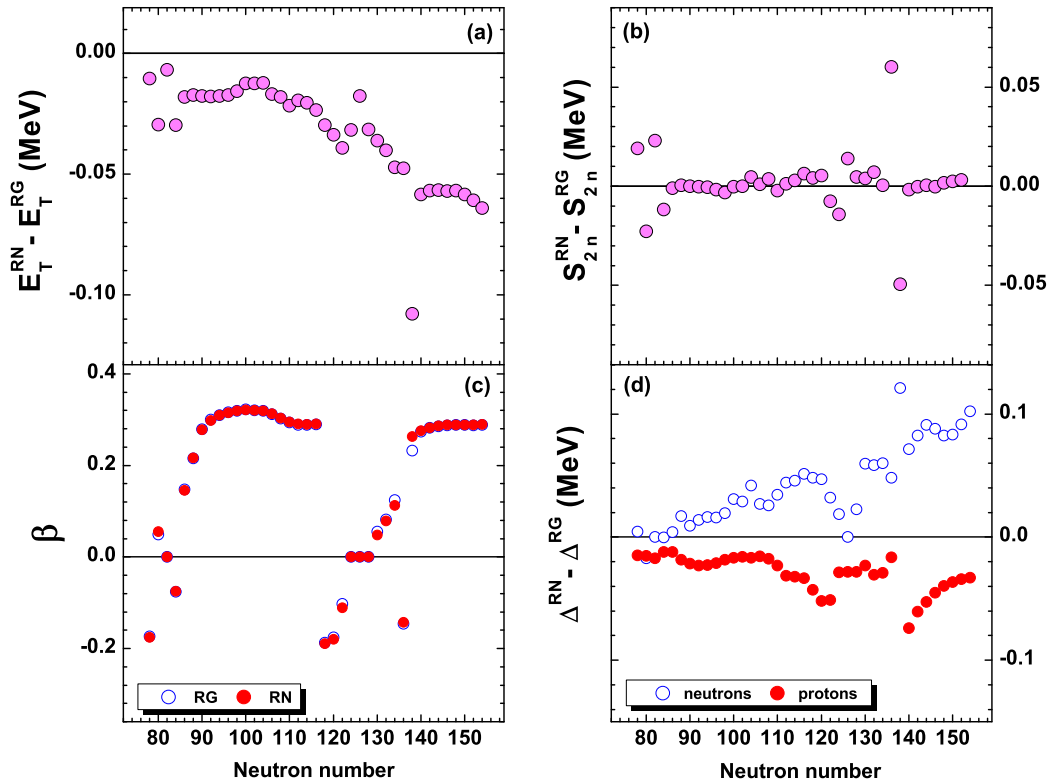}
\protect\caption{\label{regul}Differences between pairing renormalization (RN)
and regularization (RG) procedures for (a) total binding
energies, (b) two neutron separation energies, and (d) the average
neutron and proton gaps. Equilibrium quadrupole
 deformations  are shown on panel (c). Calculations are
performed for the chain of Er isotopes within the deformed
HFB+THO method
using SkP Skyrme
parametrization and mixed delta pairing.}
\end{figure}

Recently, the subject  of the  contact pairing force has been addressed in
Refs.\ \cite{[Bul02],[Bul02a],[Yu-03],[Bul04],[Nik05]} suggesting the
renormalization procedure can be replaced by a
regularization scheme which removes the cut-off energy
dependence of the pairing strength.
Differences between the HFB results emerging from the pairing
renormalization and pairing regularization procedures have
been  analyzed in Ref.~\cite{[Bor06]}
for both spherical and deformed
nuclei.
Figure.~\ref{regul}  shows  differences between the HFB-SkP
results for the deformed Er nuclei obtained using
 pairing renormalization
and regularization.
While the regularization method is better
theoretically  motivated,
it is seen that both methods give indeed  very similar  results.

\subsection{Particle Number Projection}\label{pnp}

The advantage of the mean-field approach to the pairing problem lies in
its simplicity that allows  a straightforward interpretation in terms of
pairing fields and deformations (pairing gaps) associated with the
spontaneous breaking of gauge symmetry. However, in the intrinsic-system
description, the particle-number invariance is internally broken.
Therefore, to relate to experiment, the particle number symmetry needs
to be restored. This can be done on various levels, including the
quasiparticle random phase approximation, Lipkin-Nogami (LN) method, the
projected LN method (PLN) \cite{[Dob93],[Sto03],[Sam04]}, and the
particle-number  projection before variation (PNP)
\cite{[She00a],[She02],[Ang01]}.

\begin{figure}
\includegraphics[width=0.9\textwidth]{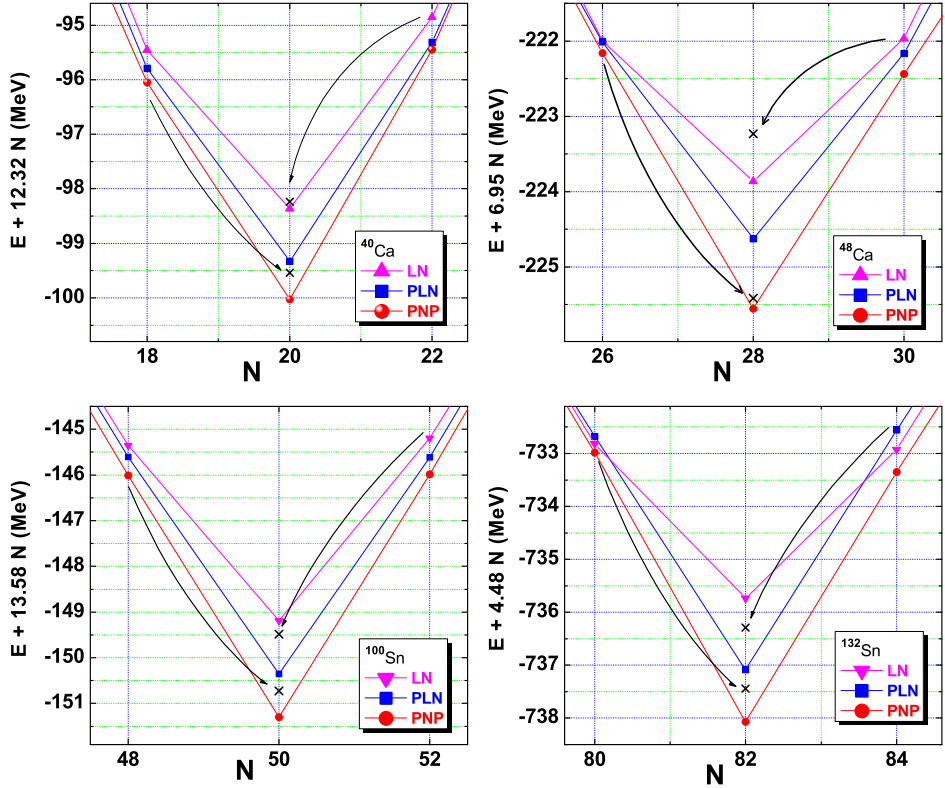}
\protect\caption{\label{proj} The total binding energy (with respect to a linear
reference) as a function of the neutron number $N$ for even-even nuclei
around doubly magic $^{40}$Ca, $^{48}$Ca, $^{100}$Sn, $^{132}$Sn calculated in LN,
PLN, and PNP methods. The crosses for magic nuclei  indicate the PLN
results  obtained by projecting from neighboring nuclei, as indicated by
arrows.}
\end{figure}

Recently, particle-number restoration before variation has been
incorporated  for the first time into the Skyrme-DFT framework employing
zero-range  pairing \cite{[Sto05a]}. It was demonstrated that the
resulting projected HFB equations can be expressed in terms of local
gauge-angle-dependent densities. In Ref.~\cite{[Sto05a]}, results of PNP
calculations have been compared with those obtained within LN and PLN
methods. While the  PLN  gives results close to PNP for  open-shell
nuclei, for   closed-shell nuclei it  breaks down with more than one MeV
difference in the total binding energy; see Fig.~\ref{proj}. This
pathological behavior of LN and PLN methods around closed-shell nuclei
can be partly cured by performing particle-number projection from
neighboring open-shell systems \cite{[Mag93]}. This result is 
important in the context
of large-scale microscopic mass calculations such as those of
Ref.~\cite{[Sam04]}. To be on the safe side, however, it is always
recommended to apply the complete PNP procedure around closed shells.

\section{Towards the Universal Nuclear Energy Density Functional}
\label{sect5}

Developing a  nuclear density functional  requires a better
understanding of the density and gradient dependence, spin and isospin
effects, and pairing, as well as an improved treatment of 
symmetry-breaking effects and many-body correlations. 
Below are summarized  the
areas of current theoretical activities in this field.

\subsection{Density and gradient dependence}

An important avenue  is
to enrich the density dependence of the isoscalar and isovector
coupling constants, both in the p-h \cite{[Dug03],[Coc04]}
and p-p channels \cite{[Dob01a],[Dob02c],[Dug04]}.
In particular, as the energy functional  is supposed to
describe those nuclear features that are related to collective dynamics,
it seems  important to
enrich the density dependence of the effective
mass in order to differentiate between its value in the bulk and at
the Fermi surface \cite{[Far01],[Gor03]}.

One of the crucial challenges in microscopic theory of nuclear
masses is to better understand  salient features of the nuclear
symmetry energy. The symmetry energy can be extracted directly from
the calculated binding energy of finite  nuclei,
after subtracting shell effects \cite{[Rei05]}. The goal is to
understand connections between the symmetry energy and isoscalar and
isovector mean fields, and in particular the influence of effective
mass and pair correlations on symmetry energy versus the isospin.
Such understanding will allow us to better determine isospin
corrections to nuclear mean fields and energy density functionals.

Recently, important indications on how to construct the nuclear
energy functional have been obtained within the  effective
field theory (see, e.g., Refs.\cite{[Pug03],[Bha04],[Fur05]}).
 Even if one still has to readjust and
fine-tune the parameters for a precise description of nuclear data,
one can gain  important insights into the structure of the
functional, especially the dependence of the coupling constants on
nuclear densities.
In addition, the  systematic, controlled  momentum expansion on which
the  effective field theory is based
offers  a way to estimate  theoretical errors (see Sec.~\ref{errors}).

\subsection{Time-odd fields}

In the self-consistent  method, the average nucleonic field is obtained
from the nucleonic density. Consequently, in a highly polarized
high-spin state,  the mean-field  potential is expected to acquire
appreciable time-odd components \cite{[Dob95],[Afa00]}. However, such
terms  should  be present in all nuclear states with non-zero angular
momentum, including ground states of odd-mass and odd-odd nuclei
\cite{[Ben00d]}. It is rather clear that without getting a handle on the
time-odd fields, it will be impossible to make precise predictions for
binding energies of most of the nuclei.

The  time-odd terms are very poorly known. An important task is 
to learn
about them through  an analysis of high-spin states and spin-isospin
excitations. Some of the time-odd fields have been studied in
Ref.~\cite{[Ben02]} in the context of Gamow-Teller beta decays in
radioactive nuclei by constraining the energy functional to the
empirical spin-isospin Landau  parameters.  The coupling constants of
the remaining terms can, in principle, be found by performing systematic
studies of rotating nuclei. This strategy has recently been followed in
the Skyrme-HF analysis of high-spin  terminating states
\cite{[Zdu05x],[Sat05b]}. Those fully aligned states have fairly simple
single-particle  configurations, and they provide an excellent  testing
ground for the time-odd densities and fields.

\subsection{Dynamical corrections}

The correlation term, accounting  for correlations going beyond the
simple product state, is an integral part of the DFT. Since nuclei are
self-bound systems, many-body correlations due to spontaneous 
symmetry-breaking effects are of particular importance. A large part of those
correlations can indeed be included by considering symmetry-breaking
product states. Within the mean-field approach, one can understand many
physical observables by directly employing broken-symmetry states;
however, for finite systems, a quantitative description often does require
symmetry restoration. For this purpose, one can apply a variety of
theoretical techniques, in particular projection methods,  the generator
coordinate method (GCM), the random phase approximation (RPA), and various
approximations performed on top of self-consistent mean fields
\cite{RS80,ReiG87,[Ben06]}.

In this context, it is important to recall that the realistic energy
density functional does not have to be related to any given effective
Hamiltonian. This creates a problem if a symmetry is spontaneously
broken. While the projection can be carried out in a straightforward
manner for energy functionals that  are related to a Hamiltonian, the
restoration of  spontaneously broken symmetries of a general density
functional   still poses a  conceptional dilemma that needs to be properly
addressed \cite{[Per95],Sto05int,BenDugint}.

Since the correlation term is a part of the functional, it should be
treated as such  during the variational procedure and during the fitting
process in which the functional's coupling constants  are determined. So
far, perhaps with the exception of the center-of-mass term (see
Sec.~\ref{com} below), such an ambitious program  has not been  carried
out. In the near future, one hopes to  work out approximate
expressions for the correlation term that would capture the essence of
results of microscopic calculations performed on top of
self-consistent mean fields. In this way, the hope is to develop the
tractable parametrization of the correlation energy in terms of local
densities that would allow  an explicit inclusion of dynamical effects
into the energy functional.

\subsubsection{Center-of-mass correction}\label{com}

The center-of-mass (c.m.)  correction, due to the violation of the
translational invariance, is always included in calculations, but its
practical implementations differ  from functional to functional
\cite{[Ben00d],[Ben03]}. For some functionals the treatment is fully
variational; for some others the c.m. term is computed following the HFB
procedure; for some functionals  a simple one-body approximation is
used. These apparently technical differences do matter as the actual
form of the c.m.  correction has a significant impact on the surface
properties \cite{[Ben00d]}.

A good  example nicely illustrating the above  point has recently been
discussed in  Ref.~\cite{[Rei05]}: for  the two functionals, SLy4 and
SLy6, which were fitted with precisely the same strategy but differ in
their treatment of c.m. correction, the surface energy coefficient
differs by as much as 0.7 MeV. While the two-body (albeit perturbative)
treatment of the c.m. correction does not reduce the overall rms error
of the fit to nuclear masses \cite{[Gor03]}, it certainly has a
significant impact on binding energies of highly deformed configurations
(such as fission isomers), fission barriers, and fission trajectories.

\subsubsection{Particle-number and isospin corrections}

As already discussed in Sec.~\ref{pnp}, efficient numerical codes that
allow for large-scale, self-consistent variational calculations after
projecting onto a good particle number have been developed
\cite{[Sto05a]}. The particle-number conserving HFB equations
\cite{[She00a],[She02]}  with Skyrme functionals can be simply obtained
from  the standard Skyrme-HFB equations in coordinate space by replacing
the intrinsic densities and currents by their gauge-angle dependent
counterparts. Using the VAP method, one can properly describe
transitions between normal and superconducting phases in finite systems,
which are inherent in (semi)magic nuclei.

As mentioned above, the restoration of broken symmetries in the
framework of DFT  causes a number of  questions, mainly related to the
density dependence of the underlying interaction and to different
treatment of particle-hole and particle-particle channels
\cite{[Ang01],[Sto04P]}. These questions are a matter of ongoing
intensive research \cite{Sto05int,BenDugint}.

Related to the particle-number symmetry, but different in origin and
treatment,  is the question of the spontaneous isospin breaking. The
isospin-breaking correction is of particular importance around the
$N$$\sim$$Z$ line. The isoscalar  pairing is believed to contribute to
the additional binding of $N$=$Z$ nuclei, the so-called Wigner energy
\cite{Sat05enam}. However, basic questions regarding the collectivity of
such a phase still remain unanswered, and should be part of the future
scientific agenda.

Apart from the presence of  charge-dependent terms in the functional,
such as the Coulomb term, the isospin symmetry is broken by  the
quasiparticle mean field (the generalized product wave function is not
an eigenstate of isospin). Several techniques have been developed to
restore isospin (see the discussion  in Refs.~\cite{[Per03],[Glo04]}  and
references quoted therein). It is fair to say, however, that in spite of
many  attempts to extend the quasiparticle approach  to incorporate the
effect of proton-neutron correlations, no symmetry-unrestricted
mean-field calculations  of proton-neutron pairing, based on realistic
effective interaction and the isospin-conserving formalism, have been
carried out so far.

\subsubsection{Rotational and vibrational zero-energy corrections}

The rotational-vibrational correlations are  important aspects
of  nuclear  collective dynamics;
they also  contribute to  nuclear binding through
quantum zero-point corrections. To estimate the magnitude of
the rotational-vibrational corrections, one usually applies
RPA \cite{[Bar04]}, GCM \cite{[Ben06]}, or
the Gaussian overlap approximation to GCM
\cite{[Ben04],[Fle04],[Pro04],[Gou05]}.

Regardless of  the  approach used, a key point
is the choice of  collective
subspace. In the case of GCM and related methods, the collective
manifold is determined by the set of external fields associated
with the  collective motion of the system.
In most practical applications, one considers five  quadrupole
degrees of freedom that give rise to nuclear rotations and
quadrupole vibrations, octupole deformations,
and pairing vibrations \cite{[Sie04],[Bar04]}.
An important step towards the microscopic description
of correlation energies are the recent large-scale benchmark calculations
of ground-state quadrupole correlations of binding energies
for all even-even nuclei, from $^{16}$O up to the superheavy
systems \cite{[Ben06]}.

\subsection{Fitting Strategy and Error Analysis}\label{errors}

One of the still-unsolved questions is an appropriate selection of
experimental data that would allow for a
more-or-less unique determination
of the coupling constants defining the energy functional. To this end,
one usually uses certain constraints obtained by extrapolating nuclear
data to an infinite system and selected data for finite nuclei. The
sensitivity of the final fit to the choice of this data set leads to a
plethora of parameterizations currently available in the literature.

Most of the currently used density functionals correctly reproduce
generic trends in nuclear masses -- as selected masses are usually
considered in the data set -- but their descriptions of other
quantities vary. Moreover, they often significantly differ in parameters
or coupling constants \cite{[Ben03]}. This suggests that yet-unresolved
correlations may exist between these parameters, and only certain
combinations thereof  are important \cite{[Ber05],[Rei05]}. Such
correlations would explain the fact that widely different
parameterizations lead to fairly similar results.

The present stage of theory requires constructing new energy density
functionals supplemented by a complete error and covariance
analysis. It is not sufficient to ``predict'' properties of exotic
nuclei by extrapolating properties of those measured in experiment.
One must also quantitatively determine errors related to such an
extrapolation. Moreover, for  experimental work it is essential
that an improvement gained by measuring one or two more isotopes be
quantitatively known. From a theoretical perspective, one must also
know the confidence level with which the parameters of the
functional are determined. An analysis of this type constitutes a
standard approach in other domains of physics, but they are seldom
performed in theoretical nuclear structure research.

\section{Summary}
\label{sect6}

This paper discusses the  status, advances, open problems, and
perspectives in the area of large-scale microscopic nuclear mass
calculations. This field of research is past phenomenological
approaches that gave us a very good understanding of general features
and trends, but lacked fundamental derivations and had limited
predictive power. At present, the focus is on microscopic
descriptions of nuclei whereupon they are treated as finite quantum
objects built of (quasi)nucleons. Nuclear ground states and masses
are in this approach determined by basic fields, which are the
particle and spin densities along with their derivatives and
gradients up to the second order in relative momenta. These fields
interact in such a way that the total energy of a given system is a
functional of densities, defined and understood in the general
framework of the Kohn-Sham theory. The determination of such a universal
functional, along with all the dynamic corrections required by data,  is
the main purpose of current investigations. In this endeavor, we
strive not only to have a reliable theoretical tool to calculate
properties of very exotic systems that will not be soon accessible in
experiment, but also wish to have a spectroscopic quality description
of well-known systems, in which very precise data do exist now, and
can be used as a rich source for determination of theoretical
parameters. Such a program of research should npt only be rooted in the
fundamentals of low-energy QCD methods and ideas, but also, by
definition, must rely on experiment for elements that cannot be
derived from first principles. It is a vast and ambitious program
presently under way.

\section*{Acknowledgments}
This work
was supported in part by the U.S.\ Department of Energy under
Contract Nos.\ DE-FG02-96ER40963 (University of Tennessee),
DE-AC05-00OR22725 with UT-Battelle, LLC (Oak Ridge National
Laboratory), by the National Nuclear Security Administration under
the Stewardship Science Academic Alliances program through DOE
Research Grant DE-FG03-03NA00083;  by the Polish Committee for
Scientific Research (KBN) under contract N0.~1~P03B~059~27 and  by
the Foundation for Polish Science (FNP).



\end{document}